\documentclass[preprint,prb,aps]{revtex4}
\usepackage{amsmath,amssymb}
\usepackage{graphicx}
\usepackage{dcolumn}
\usepackage{bm}

\begin{document}

\title{On the complete integrability of a nonlinear oscillator from group theoretical perspective}
\author{A.~Bhuvaneswari, V.~K.~Chandrasekar, M.~Senthilvelan and M.~Lakshmanan}
\affiliation{Centre for Nonlinear Dynamics, School of Physics, Bharathidasan University, Tiruchirappalli 620024, Tamil Nadu, India}

\begin{abstract}
In this paper, we investigate the integrability aspects of a physically important nonlinear oscillator which lacks sufficient number of Lie point symmetries but can be integrated by quadrature.  We explore the hidden symmetry, construct a second integral and derive the general solution of this oscillator by employing the recently introduced $\lambda$-symmetry approach and thereby establish the complete integrability of this nonlinear oscillator equation from a group theoretical perspective.
\end{abstract}


\maketitle
\section{Introduction}
\label{sec1}
In this paper, we consider the nonlinear oscillator equation,
\begin{equation}
\Delta(t, x, \dot x, \ddot x)=(1+kx^2)\ddot x-kx{\dot x}^2+\alpha^2x=0,
\label{eq1}
\end{equation}
where $k$ and $\alpha$ are arbitrary parameters and overdot denotes differentiation with respect to $t$, which was introduced by Mathews and Lakshmanan \cite{new1}.  Eq.(\ref{eq1}) is obtainable from the Lagrangian $L=\frac{1}{2}\bigg(\frac{\dot x^2-\alpha^2 x^2}{1+kx^2}\bigg)$ which can be considered as the one dimensional analogue of the Lagrangian density $\L=\frac{1}{2}\bigg(\frac{1}{1+k\phi^2}\bigg)(\partial_{ \mu} \phi \partial^{\mu} \phi-\alpha^2\phi^2)$ appearing as
a nonpolynomial model in quantum field theory \cite{new1}. Here $\alpha$ and $k$
are parameters. Eq.(\ref{eq1}) admits the general solution of the form
$x(t)=A\sin(\Omega t+\phi)$, where $A$ and $\phi$ are two arbitrary constants and the frequency
$\Omega$ is related to the amplitude $A$ through the relation $\Omega^2=\frac{\alpha^2}{1+k A^2}$.
While Mathews and Lakshmanan have also proved the quantum exact solvability of Eq.(\ref{eq1}) \cite{math1},
its three dimensional version was shown to be solvable both classically
and quantum mechanically by Lakshmanan and Eswaran \cite{lak1}. Later
Higgs and Leemon studied the classical and quantum dynamics
of the oscillator (\ref{eq1}) on the spherical configuration space \cite{higgs, leemon}.

In recent times multifaceted  investigations have been carried out on this equation
and its higher dimensional generalizations at the classical as well as quantum levels.
For example, Cari\~nena and his collaborators have generalized the above nonlinear oscillator
model to 2- and 3-dimensions and formulated an $n$-dimensional integrable analogue of it and
unearthed several mathematical properties associated with the underlying equations, including general solution,
Lagrangian and Hamiltonian structures, conserved quantities and so on \cite{nlo}.
The authors have also demonstrated that the underlying system is a superintegrable one.
A geometric interpretation of the higher dimensional system was
also proposed in relation with the dynamics
on spaces of constant curvature \cite{nlo}.
Interestingly, Eq.(\ref{eq1}) is linearizable to harmonic oscillator
through a nonlocal transformation.
The two dimensional generalization is also linearizable \cite{chan}.
By making use of the geometric properties of the kinetic energy of the corresponding Hamiltonian,
the associated quantum version of system (\ref{eq1}) was studied and considered as a deformation
of the linear harmonic oscillator \cite{san1, qn1}. The quantum version of the above nonlinear harmonic
oscillator has also been shown to be factorizable and its shape invariance property was brought
out.

Very recently Midya and Roy have generalized this nonlinear oscillator
and analyzed the underlying Schr\"{o}dinger equation under various aspects including exact solvability,
quasi-exactly solvability and
non-Hermitian invariants \cite{mid1}.
The coherent states for this nonlinear oscillator has also
been constructed and its dynamics was studied in Ref. \onlinecite{mid2}.
The studies recalled above indicate the necessity to make further
progress on the physical and mathematical structures behind the model (\ref{eq1}).

The aim of this paper is to establish the complete integrability of Eq.(\ref{eq1})
through a Lie point symmetry analysis which is a nontrivial task since the
nonlinear oscillator equation admits only the time translational symmetry.
One can deduce the first (energy) integral from the time translational symmetry
which in turn can be integrated to yield the general solution. One can also rewrite the first integral as a linear first order equation in some new variables, which can be integrated to give the second integral (see Appendix A for details).
However, we aim to explore the second symmetry which can be used to
construct the second independent integral and leads to general solution without any quadrature.

To explore the second symmetry and its associated integral we consider the recently introduced $\lambda$-symmetry method \cite{mur3,mur,murw,mur4}.  The method of finding $\lambda$-symmetries for a second-order ordinary differential
equation (ODE) has been discussed recently by Muriel and Romero \cite{mur4}. Once the $\lambda$-symmetries are found out then one can construct the integral directly from the $\lambda$-symmetry by a four step algorithm.  The underlying integrating factor can then be deduced from the integral just by differentiating the latter with respect to $\dot x$.  It has been shown that in certain specific examples$^{13}$ that eventhough the determining equations for classical Lie symmetries admit only trivial solutions, the corresponding equations for the $\lambda$-symmetries admit nontrivial solutions. On the other hand the function $\lambda$ can be chosen in such a way that the corresponding infinitesimals can be calculated.

As far as the present problem is concerned we find the $\lambda$ functions by solving the $\lambda$-invariant equation.  Once the $\lambda$ functions are determined we then proceed to construct the integrals through the prescribed algorithm \cite{mur4}. The resultant integrals provide the general solution.

The plan of the paper is as follows.  In Sec.2, we briefly recall Lie point symmetries of Eq.(\ref{eq1}).  In Sec.3,
we present the method of finding $\lambda$-symmetries and the procedure to construct the associated integral from the $\lambda$-symmetries for a given second-order ODE.  In Sec.4, we apply the procedure to Eq.(\ref{eq1}) and obtain two
$\lambda$-symmetries such that the associated first integrals are functionally independent.  From the latter we deduce the general solution.
Finally, we present our conclusions in Sec.5.

\section{Lie point symmetries of Eq.(\ref{eq1})}
In this section we study the Lie point symmetries of Eq.(\ref{eq1}) and show that it admits only time translational symmetry.  Let the evolution equation
be invariant under the one parameter Lie group of infinitesimal
transformations,
\begin{eqnarray}
&&\tilde{t}  = t+\epsilon \xi(t,x)+O(\epsilon^2), \nonumber\\
&&\tilde{x}  =  x+\epsilon \eta(t,x)+O(\epsilon^2), \quad \epsilon \ll 1,
\label{2.1}
\end{eqnarray}
where $\xi$ and $\eta$ represent the infinitesimal symmetries associated
with the variables $t$ and $x$ respectively. The associated infinitesimal
generator can be written as
\begin{eqnarray}
V  =\xi(t,x)\frac{\partial}{\partial t}+ \eta(t,x)\frac{\partial}
{\partial x}.
\label{2.2}
\end{eqnarray}
Eq. (\ref{eq1}) is invariant under the action of (\ref{2.2}) iff \cite{olv, blu1, ibr}
\begin{eqnarray}
V^{(2)}(\Delta)|_{\Delta=0} = 0,
\label{2.3}
\end{eqnarray}
where
\begin{eqnarray}
V^{(2)} = \xi\frac{\partial }{\partial t}+ \eta\frac{\partial }{\partial x}+
\eta^{(1)}\frac{\partial }{\partial\dot{x}}+
\eta^{(2)}\frac{\partial }{\partial\ddot{x}}
\label{2.4}
\end{eqnarray}
is the second prolongation in which
\begin{eqnarray}
\eta^{(1)} = \dot{\eta}-\dot{x}\dot{\xi},\;\;\;
\eta^{(2)} = \ddot{\eta}-\dot{x}\ddot{\xi}-2\ddot{x}\dot{\xi},
\label{2.5}
\end{eqnarray}
and dot denotes total differentiation.

By analyzing Eq.(\ref{eq1}) we get the following determining equations,
\begin{eqnarray}
\label{deq1}
\hspace{-1.5cm} (1+k x^2)\xi_{xx}+5kx\xi_x=0, \\
\hspace{-1.5cm} (1+k x^2)(\eta_{xx}-2\xi_{tx})+3kx\eta_x-4k x \xi_t+\eta k(1-k x^2)=0, \\
\label{deq2}
\hspace{-1.5cm} (1+kx^2)(2\eta_{tx}-\xi_{tt})+2k x \eta_t+3\alpha^2 x \xi_x=0, \\
\label{deq3}
\hspace{-1.5cm} (1+kx^2)^2\eta_{tt}+\alpha^2 x(2\xi_t-\eta_x)(1+kx^2)-\eta\alpha^2(1-kx^2)(1+kx^2)^2=0,
\label{deq4}
\end{eqnarray}
where the subscript denotes partial differentiation.  Solving Eqs. (\ref{deq1})-(\ref{deq4}), we find that the only admissible solution is $\xi=constant$ and $\eta=0$.  As a result one obtains only the time translational symmetry, that is
\begin{equation}
V=\frac{\partial}{\partial t}.
\end{equation}
Consequently one is interested to investigate the existence of more general symmetries to establish the integrability of (\ref{eq1}).

\section{$\lambda$-symmetries and integrals}
In order to be self-contained, in this section, we briefly recall the notion of $\lambda$-prolongation and the method of finding $\lambda$-symmetries and constructing the first integral from the $\lambda$-symmetry \cite{mur3, mur, murw, mur4}.

\subsection{$\lambda$-prolongation}
Let $\Delta(t,x,\dot x,\ddot x)=0$  be a second-order ODE.  A vector field $V,$ is a
symmetry of the second-order equation if there exists a function such that
\begin{equation}
V^{[\lambda,(2)]}(\Delta(t,x,\dot x,\ddot x))=0 \;\; when \;\;\Delta(t,x,\dot x,\ddot x)=0,
\label{eqnn}
\end{equation}
where $V^{[\lambda,(2)]}$ is given by
\begin{equation}
 \quad\quad
V^{[\lambda,(2)]}=\xi(t,x)\frac{\partial}{\partial t}+\eta^{[\lambda,(0)]}(t,x)\frac{\partial}{\partial x}+\eta^{[\lambda,(1)]}(t,x,\dot x)\frac{\partial}{\partial \dot x}+\eta^{[\lambda,(2)]}(t,x,\dot x,\ddot x)\frac{\partial}{\partial \ddot x},
\label{pro1}
\end{equation}
with
\begin{eqnarray}
\eta^{[\lambda,(0)]}&=&\eta(t,x),\qquad \qquad \qquad\qquad \qquad \\
\eta^{[\lambda,(1)]}&=&(D_t+\lambda)\eta^{[\lambda,(0)]}(t,x)-(D_t+\lambda)(\xi)\dot x,\\
\eta^{[\lambda,(2)]}&=&(D_t+\lambda)\eta^{[\lambda,(1)]}(t,x,\dot x)-(D_t+\lambda)(\xi)\ddot x.
\end{eqnarray}

It has been proved that \cite{murw} if $\tilde V=\xi(t,x)\frac{\partial}{\partial t}+\eta(t,x)\frac{\partial }{\partial x}$ is a $\lambda$- symmetry of the ODE $\ddot x=\phi(t,x,\dot x)$ for some function $\tilde\lambda=\tilde\lambda(t,x,\dot x)$ then $V=\frac{\partial}{\partial x}$ is also a $\lambda$-symmetry of the same equation for the function $\lambda=\tilde\lambda+\frac{A(Q)}{Q}$, where $Q=\eta-\xi \dot x$ is the characteristic of $\tilde V $ and
$A = \frac{\partial}{\partial t}+\dot{x} \frac{\partial}{\partial x}+\phi(t, x, \dot{x})\frac{\partial}{\partial \dot{x}}$ is the
vector field associated with the equation under consideration \cite{murw}. In this case the $\lambda$ determining  equation simplifies to
\begin{equation}
\phi_x+\lambda \phi_{\dot x}= \lambda_t + \dot{x} \lambda_x + \phi \lambda_{\dot{x}}+\lambda^2.
\label{fin1}
\end{equation}

In the following we solve the invariant condition (\ref{fin1}) and obtain the $\lambda$ symmetry.

We have already observed that Eq.(\ref{eq1}) admits the infinitesimals
$\xi=constant$, $\eta=0$.  Let us consider $\tilde V=\partial_t$ is a $\lambda$-symmetry of (\ref{eq1})
with $\tilde\lambda=0$.  In this case $Q=-\dot x$, ${\displaystyle\lambda= \frac{\phi(x,\dot x)}{\dot x}}$, where
${\displaystyle \phi(x, \dot{x}) = \frac{x (k \dot{x}^2 - \alpha^2)}{(1 + k x^2)}}$.
Now one can directly check that this $\lambda$ is a solution of Eq.(\ref{fin1}) as well.

If we manage to obtain second solution, $\lambda_2$, which is different from $\lambda_1$
then we can construct the second integral from $\lambda_2$ by using the same algorithm given below. In this
case $V = \frac{\partial}{\partial x}$ is both a $\lambda_1$-symmetry
and a $\lambda_2$-symmetry \cite{mur4}. Since the functions $\lambda_1$ and $\lambda_2$ are different the associated integrals are also independent.

\subsection{First integrals}

The method of generating the first integral from the $\lambda$- symmetry essentially consists of solving the following two equations (see Ref.[15]), that is
\begin{subequations}
\begin{eqnarray}
v^{[\lambda,(1)]}I&=&I_x+\lambda I_{\dot{x}}=0,\\
D_tI&=&0,\label{p1}
\end{eqnarray}
\end{subequations}
where $v^{[\lambda,(1)]}$ is the first-order $\lambda$-prolongation of $v$ and $D_t=\frac{\partial} {\partial t}+\dot{x}\frac{\partial} {\partial x}+\phi (t,x,\dot{x})\frac{\partial} {\partial \dot{x}}$ is the total differential operator.
To begin with let us suppose that $w(t,x,\dot{x})$ is a nontrivial integral $I$ of $v^{[\lambda,(1)]}$, besides the trivial integral $I=t$ (or a function of $t$), that is $w(t,x,\dot{x})$ is a solution of the first-order partial differential equation
\begin{equation}
w_x+\lambda(t,x,\dot{x})w_{\dot{x}}=0,\label{a1}
\end{equation}
where subscripts denote partial derivative with respect to that variable.
\subsubsection*{\bf Step 1:}
Find a first integral $w(t,x,\dot{x})$ which is the solution of the equation (\ref{a1}). Since there is no $t$ derivative in the equation $v^{[\lambda,(1)]}I=I_x+\lambda I_{\dot{x}}=0$, the integral of $v^{[\lambda,(1)]}$ can be written in the most general form as,
\begin{equation}
I(t,x,\dot{x})=G(t,w(t,x,\dot{x})),~G_w\neq 0,
\end{equation}
for some function of two variables $G(t,w)$. Now substituting this form in the second condition (\ref{p1}), we get
\begin{equation}
D_tI=G_t+D_t(w)G_w=0\label{ee1}.
\end{equation}
In other words, $D_t(w)=-\frac{G_t} {G_w}$, which will obviously be again a function of the variables $t$ and $w$. We call this function as $F(t,w)$. Consequently, Eq.(\ref{ee1}) can be expressed as the first-order ODE
\begin{equation}
G_t+F(t,w)G_w=0.\label{eee1}
\end{equation}
Note that in the case where $w$ itself is an integral, then $D_t w=0=F(t,w)$.
\subsubsection*{\bf Step 2:}
Evaluate $D_t w$ and express $D_t w $ in terms of $(t,w)$ as $D_t w =F(t,w)$, such that $F$ is only a function of $t$ and $\omega$.
Here $D_t$ is the total differential operator.

\subsubsection*{\bf Step 3:}
Find a first integral $G(t,w)$ of $[\partial_t+F(t,w)\partial_w] G(t,w) = 0$.
\subsubsection*{\bf Step 4:}
Then the required integral is $I(t,x,\dot{x})=G(t,w(t,x,\dot{x}))$.\\

Also, $\mu(t,x,\dot{x})=I_{\dot{x}}$
is an integrating factor of the given second-order ODE.

\section{Integrability of Eq. (\ref{eq1})}

\subsection{Determination of $\lambda$-symmetries}
As we mentioned at the end of Sec.3.2,  $\lambda=\frac{\phi(x,\dot x)}{\dot x}$ is a $\lambda$-function for (\ref{eq1}).  The explicit form of this $\lambda$ is given by
\begin{equation}
\lambda_1=\frac{x}{\dot x}\bigg(\frac{k{\dot x}^2-\alpha^2}{1+kx^2}\bigg).
\label{lam1}
\end{equation}

To determine a second particular solution of (\ref{fin1}), we assume an ansatz for $\lambda$ in
the form ( a justification for this ansatz is given in Appendix A).
\begin{eqnarray}
\lambda =\frac{a_1+a_2\dot x+a_3\dot x^2+a_4\dot x^3+a_5\dot x^4}{b_1+b_2\dot x+b_3{\dot x}^2+b_4{\dot x}^3},
\label{ansatz}
\end{eqnarray}
where $a_i$ and $b_j$'s,\; $i=1,2,3,4,5$, $j=1,2,3,4$, are arbitrary functions of $t$ and $x$ and which are to be determined.
Substituting this ansatz into Eq.(\ref{fin1}) and solving the resultant equations we obtain
the second independent solution, $\lambda_2$, in the form
(the details are given in the appendix B).

\begin{equation}
\lambda_2=\frac{(\alpha^2-k\dot x^2)(\dot x+k\dot x^2 t x+k\dot x x^2+\alpha^2 k t x^3)}{(1+kx^2)(\alpha^2 x-k(\dot x^3 t+\alpha^2 \dot x t x^2-\alpha^2 x^3))}.
\label{lam2}
\end{equation}

Using this second $\lambda$-symmetry we determine the second integral.
We note here that $\lambda_1$ can also be determined using the above ansatz.
Since it has already been found out we do not give the explicit form a second time.

\subsection{Determination of Integrals}
\subsubsection{Integral from $\lambda_1$}
\subsubsection*{\bf Step 1:}
To begin with let us consider the function $\lambda_1$ given in (\ref{lam1}).
Substituting the latter in (19) we get
\begin{equation}
w_x+\frac{x}{\dot x}\bigg(\frac{k{\dot x}^2-\alpha^2}{1+kx^2}\bigg) w_{\dot{x}} = 0.
\label{lam2eq1}
\end{equation}
We solve the characteristic equation associated with (\ref{lam2eq1}),
\begin{equation}
\frac{dx}{1} = \frac{d\dot x}{\frac{x}{\dot x}\bigg(\frac{k{\dot x}^2-\alpha^2}{1+kx^2}\bigg)},
\label{lam2eq2}
\end{equation}
and obtain a solution $w$ of the form
\begin{equation}
w(x,\dot{x}) = \frac{k\dot x^2-\alpha^2}{1+k x^2}.
\label{lam2eq3}
\end{equation}
\subsubsection*{\bf Step 2:}
Now evaluating the expression $D_t w$ in terms of the variables $t$ and $w$ we find
\begin{eqnarray}
 D_t w =F(t,w)=0.
 \label{lam2eq4}
\end{eqnarray}
\subsubsection*{\bf Step 3:}
Since $F=0$, we find $\frac{\partial G}{\partial t}$=0 which in turn yields $G(t,w(x,\dot x))=G(w(x,\dot x))$.
\subsubsection*{\bf Step 4:}
Since $G$ turns out to be a function of $w$ only, we simply choose $G=w(x,\dot x)$ and obtain the integral as
\begin{equation}
I_1=\frac{k\dot x^2-\alpha^2}{1+k x^2}.
\label{i2}
\end{equation}

\subsubsection{Second integral from $\lambda_2$}
\subsubsection*{\bf Step 1:}
Let us now consider the function $\lambda_2$.
Substituting (\ref{lam2}) in (19) we get a complicated first-order linear partial differential
equation (PDE) of the form
\begin{equation}
w_x+\frac{(\alpha^2-k\dot x^2)(\dot x+k\dot x^2 t x+k\dot x x^2+\alpha^2 k t x^3)}{(1+kx^2)(\alpha^2 x-k(\dot x^3 t+\alpha^2 \dot x t x^2-\alpha^2 x^3))} w_{\dot{x}} = 0.
\label{lam2ee1}
\end{equation}
To construct a solution of the above PDE, we solve the characteristic equation associated with (\ref{lam2ee1}), namely
\begin{equation}
\frac{dx}{1} = \frac{{{(1+kx^2)(\alpha^2 x-k(\dot x^3 t+\alpha^2 \dot x t x^2-\alpha^2 x^3))}d\dot x}}{(\alpha^2-k\dot x^2)(\dot x+k\dot x^2 t x+k\dot x x^2+\alpha^2 k t x^3)},
\label{lam2eq2}
\end{equation}
and obtain a particular solution to the PDE (\ref{lam2ee1}) in  the form
\begin{equation}
w(x,\dot{x}) = \tan^{-1}(\frac{\sqrt{\frac{\alpha^2-k\dot x^2}{1+kx^2}} x}{\dot x})-\sqrt{\frac{\alpha^2-k\dot x^2}{1+kx^2}} t.
\label{lam2eq3}
\end{equation}
  \subsubsection*{\bf Step 2:}
Now we evaluate $D_t(w)$ in terms of the variables $t$ and $w$.  Here also we find that the total differential of the function becomes null, that is $D_t(w)=0$.  This in turn fixes the function $F(t,w)=0$.
 \subsubsection*{\bf Step 3:}
As a consequence one essentially obtains a trivial equation in the third step, namely $G_t=0$.
 \subsubsection*{\bf Step 4:}
Now restricting $G(w(x,\dot x))$=$w(x,\dot x)$ one obtains an explicit expression for the second integral in the form
\begin{equation}
I_2=\tan^{-1}(\frac{\sqrt{\frac{\alpha^2-k\dot x^2}{1+kx^2}} x}{\dot x})-\sqrt{\frac{\alpha^2-k\dot x^2}{1+kx^2}} t.
\label{i2}
\end{equation}

The second integrating factor can be deduced from (\ref{i2}) just by differentiating (\ref{i2})
with respect to $\dot{x}$. 

It is also of interest to note that the second integral $I_2$ given above can be determined from the first integral $I_1$ itself, once the later is known explicitly, by the procedure pointed out earlier by three of the present authors in Ref.[7]. This procedure also gives a natural justification for the ansatz (25). The details are given in Appendix A.

\subsection{General Solution}
From $I_1$ and $I_2$, we deduce the general solution of (\ref{eq1}) straightforwardly as
\begin{equation}
\qquad \qquad \quad x(t)= A \sin(\Omega t + \delta),\;\; \Omega=\sqrt{\frac{\alpha^2}{1+k I_1}}, \;\; A = \sqrt{I_1},\;\; \delta = I_2,
\end{equation}
which is the well known solution to (\ref{eq1}).
Note that this integral is related to the standard integral \cite{new1}, $I_1 = \frac{\dot{x}^2 + \alpha^{2} x^2}{(1 + \lambda x^2)}$,
through the relation $\hat{I}_1 = \frac{(I_1 + \alpha^2)}{k}$.

\section{Conclusion}
In this paper we have established the complete integrability of a widely studied nonlinear oscillator equation  through group theoretical method.  The need for this task came from the fact that the conventional Lie point symmetry analysis proceduces only one symmetry generator.  To achieve our goal we have considered the $\lambda$-symmetry approach.  The second $\lambda$-symmetry turns out to be a rational function in $x$ and $\dot x$.  The associated integral also turns out to be a complicated rational function.  From these two integrals we derived the general solution straightforwardly.  The integrating factors for the Eq. (\ref{eq1}) have also been reported.  As a by-product of this work we have also shown the utility of the $\lambda$-symmetry approach in solving nonlinear ODEs, where sufficient number of Lie point symmetries do not exist.

\appendix
\begin{appendix}
\section{A direct method of finding $I_2$ from $I_1$}
The second integral $I_2$ given in (34) can also be obtained from a knowledge of the integral $I_1$ given in (30), as follows [18].

Let us rewrite the first integral (30) in terms of new variables $w$ and $z$ in the form
\begin{equation}
I_1=\frac{(\alpha^2-k\dot{x}^2)} {2kx\dot{x}}\frac{d} {dt}(\log (1+kx^2))=\frac{dw} {dz},
\end{equation}
where 
\begin{equation}
w=\log (1+kx^2),~ z=\int{\frac{2kx\dot{x}} {(\alpha^2-k\dot{x}^2)}dt}.\label{f1}
\end{equation}
Let us consider a particular solution of (A1) as
\begin{equation}
w=I_1z,
\end{equation}
where $I_1$ is the integration constant. Now substituting the expression (A2) in (A3), we find
\begin{equation}
1+kx^2=e^{I_1z}.
\end{equation}
Using the first integral (30) we can express $\dot{x}$ in terms of $I$ and $x$, that is
\begin{equation}
\dot{x}^2=(\alpha^2-I_1(1+kx^2))/k.
\end{equation}
Substituting the expressions $x$ and $\dot{x}$ in the second equation in (A2) and differentiating and reexpressing the variables $x$ and $\dot{x}$ in terms of $t$ and $z$, we arrive at 
\begin{equation}
\frac {dz} {dt}=\frac{2} {I_1}((\alpha^2+I_1)e^{-I_1z}-I_1-\alpha^2e^{-2I_1z})^{1/2}.
\end{equation}
Integrating the above equation, we get
\begin{equation}
\hat{I}_2+t=\frac{1} {\sqrt{I_1}}\tan^{-1}\bigg(\frac{\sqrt I_1(e^{I_1z}-1)} {((\alpha^2+I_1)e^{I_1z}-I_1e^{2I_1z}-\alpha^2)^{(1/2)}} \bigg).
\end{equation}
Eq.(A7) can be simplified to yield 
\begin{equation}
I_2=\hat{I}_2\sqrt I_1=-\sqrt I_1t+\tan^{-1}\bigg(\frac{\sqrt I_1 x} {\dot{x}}\bigg).
\end{equation}
Substituting the expression (30) for the first integral $I_1$ in (A8) we get the explicit form of $I_2$ which coincides exactly with the one given in (34). Once $I_2$ is known the associated $\lambda$-symmetry can be found through the expression [15]
\begin{equation}
\lambda_2=-\frac{I_{2x}} {I_{2\dot{x}}},
\end{equation}
where $I_2$ is given in (34). Substituting the relevant derivatives in (A9) we find that $\lambda_2$ should be in rational form. To explore it systematically within the group theoretical framework, we assume a rather general form for $\lambda_2$ (vide Eq.(20)) and solve the determining equations which exactly leads to the desired expression.

\section{Method of finding $\lambda_2$}

Substituting the ansatz (\ref{ansatz}) in (\ref{fin1}) and rearranging the
resultant equation one gets a polynomial equation in $\dot{x}$ with coefficients in $a_i$ and $b_j's, \;i=1,2,3,4,5,\; j=1,2,3,4$.
Equating the coefficients of different powers of $\dot{x}$ to zero one obtains the following set of
PDEs for $a_i$ and $b_j's,$ namely

\begin{eqnarray}
(1+kx^2)^2\big(a_{5}(a_{5}-b_{4x})(1+kx^2)+b_{4}(a_{5x}(1+kx^2)+a_{5}kx)-2b_{4}a_{5}kx\big)\nonumber\\\hspace{8.3cm}-b_{4}^2k(1-k^2x^4)=0\quad
\end{eqnarray}
\begin{eqnarray}
(1+kx^2)^2\big(a_{5}((2a_{4}-b_{4t}-b_{3x})(1+kx^2)+2b_{3}kx)+b_{4}(a_{5t}+a_{4x})(1+\nonumber\\\hspace{0.4cm}
kx^2)-a_{4}b_{4x}(1+kx^2)+a_{5x}b_{3}(1+kx^2)-2kx(b_{4}a_{4}+b_{3}a_{5})\big)\nonumber\\\hspace{7.75cm}-2b_{3}b_{4}k(1-k^2x^4)=0\quad
\end{eqnarray}
\begin{eqnarray}
(1+kx^2)^2\big(a_{5}((-b_{3t}-b_{2x}+2a_{3})(1+kx^2)-b_{4}\alpha^{2}x+3b_{2}kx)+a_{4}(-b_{4t}\nonumber\\\hspace{-1.6cm}-b_{3x}
+a_{4})(1+kx^2)-a_{3}b_{4x}(1+kx^2)+b_{4}((a_{4t}+a_{3x})(1+kx^2)-a_{3}kx)\nonumber\\\hspace{-1.3cm}+b_{3}((a_{5t}+a_{4x})(1+kx^2)+a_{4}kx)
+b_{2}a_{5x}(1+kx^2)-2kx(b_{4}a_{3}+b_{3}a_{4}\nonumber \\\hspace{3.5cm}+b_{2}a_{5})\big)-(1-k^2x^4)(b_{3}^2k-b_{4}^2\alpha^{2}+2b_{2}b_{4}k)=0\quad
 \end{eqnarray}
\begin{eqnarray}
(1+kx^2)^2\big(a_{5}((-b_{2t}-b_{1x}+2a_{2})(1+kx^2)-2b_{3}\alpha^{2}x+4b_{1}kx)+a_{4}((-b_{3t}\nonumber\\\hspace{-1.2cm}-b_{2x}+2a_{3})(1+kx^2)+2b_{2}kx)+a_{3}(-b_{4t}-b_{3x})(1+kx^2)-a_{2}b_{4x}(1+\nonumber \\\hspace{-1.0cm}kx^2)+b_{4}((a_{2x}+a_{3t})(1+kx^2)
-2a_{2}kx)+b_{3}(a_{4t}+a_{3x})(1+kx^2)+b_{2}\nonumber \\\hspace{-0.1cm}(a_{4x}+a_{5t})(1+kx^2)+b_{1}a_{5x}(1+kx^2)-2kx(b_{4}a_{2}+b_{3}a_{3}
+b_{2}a_{4}\big)\nonumber \\\hspace{3.2cm}+b_{1}a_{5})-2(1-k^2x^4)(b_{2}b_{3}k+b_{1}b_{4}k-b_{3}b_{4}\alpha^2)=0\quad
 \end{eqnarray}
\begin{eqnarray}
(1+kx^2)^2\big(a_{5}((-b_{1t}+2a_{1})(1+kx^2)-3b_{2}\alpha^{2}x)+a_{4}((-b_{2t}-b_{1x}+2a_{2})\nonumber \\\hspace{-1.3cm}
(1+kx^2)+3b_{1}kx-b_{3}\alpha^{2}x)+a_{3}((-b_{3t}-b_{2x}+a_{3})(1+kx^2)+b_{2}kx\nonumber \\\hspace{-1.2cm}
+b_{4}\alpha^{2}x)+a_{2}((b_{3x}-b_{4t})(1+kx^2)-b_{3}kx)-a_{1}b_{4x}(1+kx^2)+b_{4}((a_{1x}\nonumber\\\hspace{-1.2cm}
+a_{2t})(1+kx^2)-3a_{1}kx)+b_{3}(a_{3t}+a_{2x})(1+kx^2)+b_{2}(a_{4t}+a_{3x})(1+\nonumber\\\hspace{-0.2cm}
kx^2)+b_{1}(a_{5t}+a_{4x})(1+kx^2)-2kx(b_{4}a_{1}+b_{3}a_{2}+b_{2}a_{3}+b_{1}a_{4})\big)\nonumber\\\hspace{3.3cm}
-(1-k^2x^4)(b_{2}^{2}k+2b_{1}b_{3}k-b_{3}^{2}\alpha^2-2b_{2}b_{4}\alpha^{2})=0\quad
 \end{eqnarray}
\begin{eqnarray}
(1+kx^2)^2\big(a_{4}((-b_{1t}+2a_{1})(1+kx^2)-2b_{2}\alpha^{2}x)+a_{3}((-b_{2t}-b_{1x}+2a_{2})\nonumber\\\hspace{-1.1cm}
(1+kx^2)+2b_{1}kx)+a_{2}((-b_{3t}-b_{2x})(1+kx^2)+2b_{4}\alpha^{2}x)+a_{1}((-b_{3x}\nonumber\\\hspace{-0.7cm}
-b_{4t})(1+kx^2)-2b_{3}kx)+b_{4}a_{1t}(1+kx^2)+b_{3}(a_{2t}+a_{1x})(1+kx^2)\nonumber\\\hspace{-0.8cm}
+b_{2}(a_{3t}+a_{2x})(1+kx^2)+b_{1}((a_{4t}+a_{3x})(1+kx^2)-4a_{5}\alpha^{2}x)-2kx\nonumber\\\hspace{0.45cm}
(b_{3}a_{1}+b_{2}a_{2}+b_{1}a_{3})\big)-2(1-k^2x^4)(b_{1}b_{2}k-b_{2}b_{3}\alpha^{2}-b_{1}b_{4}\alpha^2)=0\quad
\end{eqnarray}
\begin{eqnarray}
(1+kx^2)^2\big(a_{1}((-b_{2x}-b_{3t}+2a_{3})(1+kx^2)+3b_{4}\alpha^{2}x-b_{2}kx)+a_{2}((-b_{1x}\nonumber\\\hspace{-1.3cm}
-b_{2t}+a_{2})(1+kx^2)+b_{3}\alpha^{2}x+b_{1}kx)-a_{3}b_{1t}(1+kx^2)+b_{1}((a_{2x}+a_{3t})\nonumber\\\hspace{-1.0cm}
(1+kx^2)-3a_{4}\alpha^{2}x)+b_{2}((a_{1x}+a_{2t})(1+kx^2)-a_{3}\alpha^{2}x)+b_{3}a_{1t}(1+\nonumber\\\hspace{0.3cm}
kx^2)-2kx(b_{2}a_{1}+b_{1}a_{2})\big)-(1-k^2x^4)(kb_{1}^2-b_{2}^2\alpha^2
-2b_{1}b_{3}\alpha^2)=0\quad
\end{eqnarray}
\begin{eqnarray}
(1+kx^2)^2\big(a_{1}((-b_{1x}-b_{2t})(1+kx^2)+2b_{3}\alpha^{2}x)+a_{2}(2a_{1}-b_{1t})(1+kx^2)\nonumber\\\hspace{-0.4cm}
+b_{1}((a_{1x}+a_{2t})(1+kx^2)-2a_{3}\alpha^{2}x)+b_{2}a_{1t}(1+kx^2)-2kxb_{1}a_{1}\big)\nonumber\\\hspace{7.5cm}
+2(1-k^2x^4) b_{1}b_{2}\alpha^2=0\quad
\end{eqnarray}
\begin{eqnarray}
(1+kx^2)^2\big(a_{1}((-b_{1t}+a_{1})(1+kx^2)+b_{2}\alpha^{2}x)+b_{1}(a_{1t}(1+kx^2)-a_{2}\alpha^{2}x)\big)\nonumber\\\hspace{8cm}+(1
-k^2x^4) b_{1}^2\alpha^2=0\quad
\end{eqnarray}

To solve these equations we again assume $a_i$ and $b_j$'s are quartic polynomials in $x$ with
coefficients which are unknown functions in $t$. Substituting these forms in the above equations we find each one of the
above equations reshapes to a  polynomial equation in $x$. Equating the coefficients of different powers of $x$ to
zero in each equation  one obtains a set of ODEs for the differential coefficients.
Solving them consistently one finally arrives at the solution given
in (\ref{lam2}).

\end{appendix}

\section*{Acknowledgment}
    The work of VKC and ML is supported by a DST - IRHPA research project.  The work of ML is also supported by DST - Ramanna Fellowship program and a DAE Raja Ramanna Fellowship.  The work of MS forms part of a DST sponsored research project.



\begin{thebibliography}{10}
\bibitem{new1}
P. M. Mathews and M. Lakshmanan, Q. Appl. Math. {\bf 32}, 215 (1974)

\bibitem{math1}
 P. M. Mathews and M. Lakshmanan, Nuovo cimento A {\bf 26},  299 (1975)

\bibitem{lak1}
M. Lakshmanan and K. Eswaran, J. Phys. A: Math. Gen. {\bf 8 }, 1658 (1975)

\bibitem{higgs}
P. W. Higgs, J. Phys. A: Math. Gen. {\bf 12},  309 (1979)

\bibitem{leemon}
H. I. Leemon, J. Phys. A: Math. Gen. {\bf 12}, 489 (1979)

\bibitem{nlo}
J. F. Cari\~{n}ena, M. F.Ra\~{n}ada, M. Santander and M. Senthilvelan, Nonlinearity {\bf 17},  1941 (2004)

\bibitem{chan}
V. K. Chandrasekar, M. Senthilvelan and M. Lakshmanan, Proc. Roy. Soc. Lond. Series A {\bf 465}, 2369 (2009)

\bibitem{san1}
J. F. Cari\~{n}ena, M. F. Ra\~{n}ada and M. Santander, Rep. Math. Phys. {\bf 54}, 285 (2004)

\bibitem{qn1}
J. F. Cari\~{n}ena, M. F. Ra\~{n}ada and M. Santander, Ann. Phys. {\bf 322},  484 (2007)

\bibitem{mid1}
B. Midya and B. Roy, J. Phys. A: Math. Theor. {\bf 42}, 285301 (2009)

\bibitem{mid2}
B. Midya, B. Roy and A. Biswas, Phys. Scr. {\bf 79},  065003 (2009)

\bibitem{mur3}
C. Muriel and J. L. Romero, IMA J. Appl. Math. {\bf66},  111 (2001)

\bibitem{mur}
C. Muriel and J. L. Romero, J. Lie Theory {\bf 13}, 167 (2003)

\bibitem{murw}
C. Muriel and J. L. Romero, {\it $\lambda$-symmetries on the derivation
of first integrals of ordinary differential equations
in Waves and Stability in Continuous Media (WASCOM) (Eds.)  A. M. Greco, S. Rionero and T. Ruggeri} (World Scientific, Singapore, 2009)  p.303-308

\bibitem{mur4}
C. Muriel and J. L. Romero, J. Phys.  A: Math.Theor. {\bf42},  365207 (2009)


\bibitem{blu1}
G. W. Bluman and S. Kumei, {\it Symmetries and Differential Equations} (Springer-Verlag, New York, 1989)

\bibitem{ibr}
N. H. Ibragimov, {\it Elementary Lie Group Analysis and Ordinary Differential Equations} (John Wiley and Sons, New York, 1999)

\bibitem{vkc}
V. K. Chandrasekar, M. Senthilvelan and M. Lakshmanan, J. Phys. A: Math. Gen. {\bf 39}, L69 (2006) 


\end{thebibliography}
\end{document}